\documentclass[prb,preprint,preprintnumbers,amsmath,amssymb]{revtex4}

\usepackage{graphicx}
\usepackage{dcolumn}
\usepackage{bm}
\usepackage[usenames]{color}
\usepackage{float}

\begin{document}

\def\Ef{$E_{\rm F}$}
\def\Ed{$E_{\rm D}$}
\def\Eg{$E_{\rm g}$}
\def\Efmath{E_{\rm F}}
\def\Edmath{E_{\rm D}}
\def\Egmath{E_{\rm g}}
\def\Tc{$T_{\rm C}$}
\def\kpara{{\bf k}$_\parallel$}
\def\kparamath{{\bf k}_\parallel}
\def\kperp{{\bf k}$_\perp$}
\def\invA{\AA$^{-1}$}
\def\Kbar{$\overline{\rm K}$}
\def\Kbarprime{$\overline{\rm K}^\prime$}
\def\Mbar{$\overline{\rm M}$}
\def\Gbar{$\overline{\Gamma}$}
\def\DeltaEf{$\Delta E_{\rm F}$}
\def\DeltaEfmath{\Delta E_{\rm F}}
\def\DeltaSO{$\Delta_{\rm so}$}
\def\onebyone{$1\times1$}
\def\eightbyeight{$8\times8$}
\def\ninebynine{$9\times9$}


\title{Graphene for spintronics: giant Rashba splitting due to hybridization with Au}

  \author{ D. Marchenko,$^{1}$ A. Varykhalov,$^1$ M. R. Scholz,$^1$  
   G. Bihlmayer,$^3$ E. I. Rashba$^4$, A. Rybkin,$^2$ A. M. Shikin,$^2$ 
    O. Rader$^1$}  

  \affiliation{$^1$Helmholtz-Zentrum Berlin f\"ur Materialien und Energie, 
Elektronenspeicherring BESSY II, Albert-Einstein-Str. 15, D-12489 Berlin, Germany}
 \address{$^2$Institute of Physics, St. Petersburg State University, 198504, 
  St. Petersburg, Russia }

\affiliation{$^3$Peter Gr\"unberg Institut and Institute for Advanced Simulation,
Forschungszentrum J\"ulich and JARA, D-52425 J\"ulich, Germany}

 \affiliation{$^4$Department of Physics, Harvard University, Cambridge, 
Massachussetts 02138, U. S. A. }

\maketitle

\textcolor{white}{
\section*{Abstract}
}
\addtocounter{section}{1}
 {\bf
Graphene in spintronics [\onlinecite{Yazyev}] has so far primarily meant
spin  current leads of high performance  because the intrinsic
 spin-orbit coupling of its $\pi$ electrons is very weak 
[\onlinecite{Tombros-Nature-2007,Shinjo09,Kawakami09}]. 
 If a large spin-orbit coupling could be created by a proximity effect, the material could 
also form  active elements of a spintronic device such as the 
Das-Datta spin field-effect transistor [\onlinecite{Das}], however, 
metal interfaces often compromise the band dispersion 
of massless Dirac fermions [\onlinecite{OshimaPRB92}].
Our measurements show that  Au intercalation 
at the graphene-Ni interface creates a 
 giant spin-orbit splitting ($\sim100$ meV) in the graphene Dirac cone
up to the Fermi energy.
Photoelectron spectroscopy reveals  hybridization
with Au$5d$ states as the source for the giant spin-orbit splitting. 
An {\it ab initio} model of the system shows a Rashba-split dispersion
with the analytically predicted gapless band topology 
around the Dirac point of graphene and indicates that a sharp graphene-Au
interface at equilibrium distance
will account for only $\sim 10$ meV spin-orbit splitting. 
The  {\it ab initio}  calculations suggest
an enhancement due to Au atoms that get closer to the graphene
and do not violate the sublattice 
symmetry.
 }

\textcolor{white}{
\section*{Introduction}
}
\addtocounter{section}{1}
Graphene shows fascinating electronic properties due to its  structure
consisting of two equivalent sublattices A and B that determine its band structure
with the linear dispersion near conical 
  \Kbar\ and \Kbarprime\ points in reciprocal space.  
A and B are identified as a pseudospin, and the isospin indicating \Kbar\
and \Kbarprime\ valleys is a conserved quantum number [\onlinecite{KM05}].
In addition to the   pseudo- and isospin, also the real spin is
recently being considered important in graphene 
[\onlinecite{Yazyev,KM05,Huertas09,Trauzettel,Gmitra,Trickey,Semenov,Tombros-Nature-2007,Varykhalov-PRL-2008,Shinjo09,Kawakami09}].
The main contribution to the intrinsic  spin-orbit splitting originates
from the coupling to graphene $d$ orbitals [\onlinecite{Gmitra}].
The resulting splitting is just of the order of 0.01 meV [\onlinecite{Trickey}] 
which means together with the high carrier mobilities in graphene 
a large  spin coherence length  with measured
values reaching 1.5--2 $\mu$m [\onlinecite{Tombros-Nature-2007,Shinjo09,Kawakami09}].
The theoretical predictions are even by an order of magnitude higher so 
that it is at present unclear whether these values represent 
practical limits [\onlinecite{Huertas09}]. 

We have recently measured a graphene spin-orbit splitting of 
$\Delta_{\rm so}=13$ meV [\onlinecite{Varykhalov-PRL-2008}].
This splitting occurs in the system graphene/Ni(111) after a   Au layer
with one atom thickness
  is intercalated by heating between the graphene and the Ni [\onlinecite{ShikinPRB00}].
This Au layer serves a dual purpose: on the one hand, it transforms the strongly 
bonded [\onlinecite{OshimaPRB92}] graphene monolayer on Ni(111) into an electronically 
quasifreestanding structure.
This decoupling effect is apparent from the resulting linear quasirelativistic 
$\pi$ bands of the graphene, i. e., the Dirac cone [\onlinecite{Varykhalov-PRL-2008,fussnote1}].

On the other hand, the Au interlayer produces a different type of 
the spin-orbit splitting 
through an extrinsic effect. Indeed, the 
 spin-orbit coupling   is known to be important at the surface of  Au [\onlinecite{LaShell}] 
and in Au nanowires [\onlinecite{Barke06}], and 
Au(111) is frequently considered a model for a Rashba effect 
induced by breaking up-down symmetry [\onlinecite{LaShell,Henk}]. 
For graphene with its   2-atom basis (sublattices A and B) 
and an additionally broken up-down symmetry, 
the band dispersion 
including spin-orbit coupling has been calculated in Ref. [\onlinecite{RashbaPRB09}]. 
In zero magnetic field, 
the dependence of the spin polarization on the two-dimensional electron wave vector
\kpara\ was found. 
The band topology was predicted to be similar to that of the unbiased
spinless graphene bilayer but with an additional spin texture which is tangential
to the circular constant-energy surfaces. 
The effect couples spin and pseudospin in such a way that contributions 
from both A and B sublattices 
can lead to a substantial spin interference in the photoemission 
process [\onlinecite{Kuemmeth}].

\textcolor{white}{
\section*{Experiment}
}
\addtocounter{section}{1}
In the present Communication, we show by spin- and angle-resolved photoelectron 
spectroscopy that an extrinsic spin-orbit splitting \DeltaSO\  of the order of 100 meV 
can be obtained in graphene near the Fermi energy, i. e., an enhancement 
by an  
order  
of magnitude relative to our previous finding [\onlinecite{Varykhalov-PRL-2008}].
We will show that we can very well identify the type of superstructure characterising
the graphene-Au interface in good agreement with the bulk lattice constants of
graphene, Au, and Ni. 
We determine very clearly the spin-dependent hybridization at the interface
 and its change with the superstructure. Based on {\it ab initio} calculations, the  
giant size of the spin-orbit splitting must be understood as resulting
from more dilute Au at a closer distance to the carbon atoms than in a Au monolayer, 
either  as adatoms above or below the graphene layer.  
Figure 1 shows spin- and angle-resolved photoemission measurements
of graphene/Au/Ni(111). Part (a) shows the $\pi$ band with its linear 
quasirelativistic dispersion, and 
crosses indicate where spin-resolved photoemission
spectra (b--d) have been measured. The splitting between spin-up spectra $I^\uparrow(E)$
(upward triangles) and spin-down spectra $I^\downarrow(E)$
(downward triangles) is clearly visible and 
amounts to 90 meV at $\kparamath=1.65$ \invA\ 
and $-105$ meV at $\kparamath=-1.65$ \invA.
Apparently, the sign of the splitting reverses with the sign of \kpara, as 
expected for a Rashba effect.
This giant spin splitting 
constitutes our central experimental result. The splitting exceeds also the room-temperature broadening,  and together with the fact that it extends to the Fermi energy, this makes it directly relevant for transport applications. 
In fact, transport properties of graphene with an externally produced (Rashba-type)    
spin-orbit splitting have been intensively investigated in recent years [\onlinecite{Dyrdal09,QSHE,QAHE}]
and important predictions for such a constellation of graphene have been published: 
 the spin Hall effect [\onlinecite{Dyrdal09}] and 
the quantum spin Hall effect [\onlinecite{KM05,QSHE}]
and, with an additional exchange interaction, the quantum anomalous Hall effect 
[\onlinecite{QAHE}].

To establish connection between our previous 13 meV data and the new 100 meV data, we undertook closer
inspection of the spin-resolved spectra of Fig. 1. 
They reveal that the system is inhomogeneous and the high-splitting phase ($\sim100$ meV) and the 
low-splitting phase ($\sim10$ meV) are  present simultaneously.  
Line fits and spectral decompositions (see Supplementary Material, Fig. S5) show that the 
high-splitting phase   makes up between one third and two thirds 
of the spectral weight, corresponding in the simplest picture to half of the  
 sample region of about $200 \mu{\rm m} \times 200 \mu$m that is probed simultaneously. 
The decomposition is possible due to an energy shift between the two phases 
with the low-splitting phase appearing at about 200--500 meV  higher binding energy.
 The giant size of the splitting and the distribution in spectral weight
between the high-splitting phase and the low-splitting phase ranging from 
   $1:2$ to $2:1$  have been reproduced
in several {\it in situ} preparations and at different beamlines.

We want to unveil  
the electronic origin of this strongly enhanced spin splitting  including
the reason for its nonuniformity  
and the relation to our previous results. 
Probing as well as controlling the Au  layer as the likely source of the giant
Rashba splitting are challening because of its very location underneath the graphene
layer. 
The means of control is basically through the amount of Au deposited initially (before heating)
 onto the graphene which typically exceeds the subsequently 
intercalated amount.
While graphene and Ni(111) have a lattice mismatch of only 1.2\%, the one between 
graphene and Au(111) is much larger ($\sim14$\%). Therefore, the intercalated monolayer of Au
will not reach the same atom density as the Ni. 
Probing the resulting structure is possible by scanning tunneling microscopy (STM)
 because the moir\'e effect reveals 
the superstructure through  
the beat frequency between the graphene lattice and the Au monolayer.
Moreover, the photoemission signal from Au$5d$ states  under the
graphene is very well visible.
Figures 1a and 2c show that the graphene electronic structure resembles that of freestanding
graphene. 
 The Au layer prevents the strong  carbon-Ni hybridization. 
Nevertheless, the graphene $\pi$-states show also in the quasifreestanding phase
with the Dirac cone 
 several  carbon-Au hybridization points  in the range of 4 to 6.5 eV binding energy
in Fig. 2c.
Moreover, the orange arrows in Fig. 2c indicate  further replicas  of the graphene $\pi$-band shifted to 
smaller and larger values of the  
wave vector \kpara. The shift of \kpara\ amounts to between $1/7$ and $1/9$ of the distance 
\Gbar\Kbar. The STM image in Fig. 2a reveals a moir\'e pattern 
of similar periodicity but it can  most accurately  be determined from the superstructure in the 
low-energy electron diffraction (LEED) pattern
as  a \ninebynine\ superstructure ---  in good agreement with the shift of the replica bands
in photoemission. 

Dashed lines in Fig. 2c show where we probed  the hybridization points by spin-resolved spectra
(Fig. 2d).
These spectra reveal that the giant spin splitting of the $\pi$-band smoothly merges
into an even larger spin splitting ($\sim0.6$ eV) of the Au$5d$ states.
This is strong indication that this spin-dependent C$\pi$-Au$d$ hybridization with the heavy Au 
is the source of the giant Rashba splitting in the graphene. 

Interestingly, when more Au is initially deposited, it forms after the intercalation a slightly different
\eightbyeight\ superstructure. The Supplementary Material (Fig. S1) shows that it is characterised
by an extra hybridization with other Au$d$ states in the $\pi$-band. 
This hybridization does not enhance the giant spin-orbit splitting further (Fig. S6), which 
indicates already that the differences between the \ninebynine\ and \eightbyeight\ superstructures 
do  not play a role for the giant spin-orbit splitting.

\textcolor{white}{
\section*{Theory}
}
\addtocounter{section}{1}
We want to turn to {\it ab initio} theory in order to verify the giant spin splitting. 
Modelling  the \ninebynine\ or \eightbyeight\ superstructure of graphene/Au,
which would increase   the unit cell 
 by two orders of magnitude over that of graphene, is unnecessary 
because of the similar spin-orbit splitting for the two superstructures.  
A $p(1\times1)$ structure and on-top position for graphene on 
  1 monolayer (ML) Au  has  instead been chosen   (Fig. 3a).

The distance between the graphene and Au monolayer is a parameter of enormous 
influence on the spin-orbit splitting of the
graphene. The fact that also the Fermi energy varies characteristically with this distance allows us
to connect to the experiment: The experimental Fermi level corresponds with its slight
$p$-doping (hole doping) to  3.3--3.4 \AA\ in the theoretical model, and this data
is shown in Fig. 3a. 
3.3 \AA\ is also the equilibrium distance calculated before for graphene/Au(111) [\onlinecite{Giovannetti-PRL-2008}].
According to analytical prediction, the band topology is peculiar around
the \Kbar\ point  with two pairs of bands --- 
a gapped one and a non-gapped one [\onlinecite{RashbaPRB09,Kuemmeth}].
This analytical model [\onlinecite{RashbaPRB09,Kuemmeth}] does not assume  
any specific surface configuration.  
The bands have so far been confirmed for freestanding graphene in a
supercell geometry by density functional theory calculations
for an applied field $E=4.0$ V/nm [\onlinecite{Gmitra}] but not yet for the field $E$
realistically replaced by an interface to Au. 
This was done in Fig. 3a which shows the spin-orbit split
bands at the Dirac point at \Kbar\ for the on-top geometry. 
In the inset, these {\it ab initio} results (symbols) are 
 magnified near the Dirac point and compared to the
dispersion of the analytical model (solid line), which is fully confirmed. 
This is also important for the predicted magnetic-field
dependence [\onlinecite{RashbaPRB09}].  
In agreement with the experiment, our {\it ab initio} calculation reveals C$\pi$-Au$d$ hybridization
as the source of the spin-orbit splitting in the graphene. The graphene $\pi$-band 
($p_z$ orbitals) hybridizes indeed with the deeper Au$5d$ bands
of $d_{z^2}$ and $d_{zx}$ type because of their matching symmetry. 
Despite the large distance of 3.3 \AA\
the hybridization is strong which has not been considered in the literature in connection
with graphene-noble-metal interfaces. 
This is best seen after the
 spin-orbit coupling is turned  
off leaving only gaps caused by hybridization:
we determine hybridization gaps of widths 
 $E_{g,z^2}\approx 0.8$ eV and $E_{g,yz}\approx 0.5$ eV (see Fig. S11).
Their absolute binding energies do not compare well to the experiment
revealing the limitations of the Au monolayer as model substrate. 
 This situation improves largely and the main C$\pi$-Au$d$ hybridization moves
from $\sim3$ eV to $\sim4$ eV below \Ef\ 
when the Ni substrate is included in the calculation (see Fig. S12).

 The spin-orbit splitting in Fig. 3a     is 9 meV near \Ef\ 
but increases strongly when the distance between graphene and Au
 is reduced (see Fig. S10 for the detailed behaviour with the distance). 
On the other hand,  the Dirac cone is destroyed at closer graphene-Au distance,
giving, e. g., for 2.5 \AA\ a band gap at \Kbar\ of 40 meV. 
The reason for this is the broken A-B symmetry of carbon atoms in the on-top
geometry (Fig. 3a) which, remarkably, does not manifest itself very much at larger distances.
A hollow-site geometry, instead, preserves the A-B symmetry in the graphene, see Fig. 3b. 
Consequently, an intact Dirac cone is obtained in the hollow-site geometry
also for smaller graphene-Au distances such as 2.5 \AA\  in Fig. 3b. 
We are very well able to obtain at this arbitrary interlayer distance of 2.5 \AA\   a giant spin-orbit
splitting of $\sim70$ meV. However, such close distance
costs as much as 1 eV relative to the equilibrium separation and is thus unrealistic. 
 While the giant spin-orbit splitting apparently is difficult to reproduce by 
density functional theory in equilibrium,
the intact Dirac cone is not. The intact Dirac cone is presently obtained 
with $p(1\times1)$ on-top graphene/Au  
and has also been found for $p(1\times1)$  graphene/Cu(111) where the on-top position is
 determined to be energetically favourable [\onlinecite{Giovannetti-PRL-2008}].
Relative to this $p(1\times1)$ on-top geometry which implies maximum A-B symmetry breaking, 
a moir\'e superstructure such as the $9\times9$ one necessarily breaks the A-B symmetry 
to a lesser or vanishing degree which is favourable for obtaining an intact Dirac cone. 
After having investigated  other laterally shifted positions of the $p(1\times1)$ overlayer 
and a $4\times3$ moir\'e
superstructure as well as corrugation in the graphene all yielding spin-orbit splittings of the order
of 10 meV at equilibrium graphene-Au distances, 
the conclusion is that the giant spin-orbit splitting will not be accounted
for by the sharp graphene-Au interface alone that our structural characterization 
by LEED and STM suggests as simplest case. This means that a
 model for a realistic splitting will have to include individual Au atoms which then can 
obtain a higher coordination to the carbon and due to the
resulting attraction and shorter distance exert
a stronger spin-orbit coupling on the graphene $\pi$-states. 
As seen above, critical is at small graphene-Au separations the preservation of A-B symmetry 
which leads us once again to the hollow site geometry.
Figure 3c shows this situation for graphene/0.25 ML Au
 in a $p(2\times2)$ structure with the Au in the hollow site relative to the graphene.
The vertical distance of the Au   to the graphene layer is 2.3 \AA\ which  (as distinct from the Au monolayer of Fig. 3b)  is near the
equilibrium determined by our structural optimization. Figure 3c shows that this  
structure enhances the 
spin-orbit splitting to values between 50 and 100 meV while keeping the characteristic band topology
and the Dirac point. This demonstates that our experimental results are also plausible theoretically.

The \ninebynine\ and \eightbyeight\ structures will include in the simplest case a substantial amount 
of Au atoms arranged in the hollow sites of graphene [\onlinecite{VarykhalovPRB10}] 
but this does not imply a reduced distance, especially if one considers that \ninebynine\ is also
the structure which a Au monolayer alone forms on Ni(111) [\onlinecite{Jac95}]. 
Therefore, the   \ninebynine\ and \eightbyeight\ superstructures 
are probably not relevant for the giant splitting and both give rise to only $\sim10$ meV splitting. 

The previously measured data [\onlinecite{Varykhalov-PRL-2008}] were characterized by a smaller
$\sim 13$ meV spin-orbit splitting, and this low-splitting phase is principally in agreement with our present
calculations for the full Au monolayer. The published [\onlinecite{Varykhalov-PRL-2008}] band dispersion  
measured along \Gbar\Kbar\  reflects the presence of a sample with structural defects. 
It contained substantial contributions of rotated domains visible as characteristic
\Gbar\Mbar\ dispersions appearing along \Gbar\Kbar\ (see Fig. S2) which is not the case in the 
present data even when the amount of intercalated Au is varied systematically
from zero to more than 1 ML (see Fig. S1).  
The intercalation process under the graphene, including that of Au, is a present far from understood.
The accepted main route is via defects in the graphene [\onlinecite{Tontegode}], and for large molecules
this can be confirmed by STM directly [\onlinecite{VarC60}]. 
The presence of many domain boundaries facilitates the intercalation and apparently results in the
low-splitting phase. Intercalation of Au works, however, also in samples which are free of
defects over large distances. The temperature of intercalation and that of the initial graphene formation
are very similar in the graphene/Au/Ni(111) system so that an opening and closing of the graphene appears
possible during intercalation. 
This could lead to more Au locally closer to the graphene, either as subsurface Au
or as adatoms. 

The $p(2\times2)$ plot of Fig. 3c is  useful for demonstrating that an enhancement of the 
spin-orbit coupling in graphene by    sparsely distributed Au atoms is possible in an 
equilibrium structure but it does  not imply 
that the Au arrangement  possesses 
such an ordered structure. 
We can exclude that the Au atoms substitude carbon atoms. Their size would correspond to two carbon atoms 
and the resulting distortion  would manifest  itself in STM.  
Another possibility are non-intercalated residual atoms above the graphene
for which we have no direct experimental evidence. We mention that such atoms  
cannot be detected by
our STM technique
because they should  
be highly mobile on graphene and  
shifted along by the STM tip. 
In order to explore a realistic model which includes the Ni substrate, the
 band dispersion for a structure of graphene sandwiched between 0.25 ML Au 
and 1 ML Au on top of 3 ML Ni has been calculated as well (Fig. S11) and gives practically
the same Dirac cone with giant spin splitting as in Fig. 3c.

\textcolor{white}{
\section*{Summary}
}
\addtocounter{section}{1}
In summary, we report  
a giant Rashba splitting in  
graphene in
contact with Au up to 100 meV that 
is caused by   graphene-Au hybridization. 
A flat Au monolayer can account for only $\sim10$ meV spin-orbit splitting 
as has been reported in Ref.~[\onlinecite{Varykhalov-PRL-2008}],
whereas a structure 
including laterally more separated Au adatoms residing
in  
hollow-site  
positions closer to graphene, gives at the equilibrium graphene-Au distance
rise to the enhanced splitting of $\Delta_{\rm SO} \sim 50$--$100$ meV, 
a realistic Fermi level position,
and an intact Dirac cone. We attribute simultaneous presence of both 
100 meV and 10 meV splittings to the coexistence of areas without and with 
extra Au either as adatoms or immersed into the graphene. 
 
 \phantom{x}

\textcolor{white}{
\section*{Methods}
}
\addtocounter{section}{1}
\noindent {\bf Methods}

Angle-resolved spectra have been measured with a hemispherical 
analyzer and parallel angular detection and a 6-axes automated  manipulator.
Spin- and angle-resolved photoemission has been performed with a 
hemispherical analyzer coupled to a Rice University Mott-type spin
polarimeter operated at 26 kV sensitive to the in-plane spin component
perpendicular  to \kpara. 
Linearly polarized synchrotron light from the UE112-PGM undulator 
beamlines at BESSY II has been used for excitation. 
The unmagnetized Ni(111) surface was prepared as a 15--20-ML thick film on W(110), 
and graphene was synthesized by cracking of propene at the Ni(111) 
surface held at $T \sim800$ K. 
Because the surface reactivity drops drastically with the graphene covering, this
procedure results in exactly one graphene monolayer [\onlinecite{OshimaPRB92}]. 
Intercalation of Au [\onlinecite{ShikinPRB00}] was achieved by deposition 
of a nominal monolayer coverage of Au
on the graphene followed by brief annealing   at $750$ K. 
This Au coverage was systematically varied   in a wedge-type experiment. 
Overall energy (of electrons and photons) and angular resolution  of the experiments was 
80 meV and 1$^\circ$ with both angle and spin resolution and 6 meV and 0.4$^\circ$ with
angle resolution only. 
The base pressure was $1$--$2 \times 10^{-10}$ mbar. 

 The calculations have been conducted in the generalised gradient approximation 
[\onlinecite{ref32}] to density functional theory, using the full-potential linearised
augmented planewave method  implemented in the FLEUR code [\onlinecite{ref33}].
We use a plane-wave cutoff of 7.37 \invA\ and muffin-tin radii
of 0.68 \AA\ for carbon, 1.22 \AA\ for Au and 1.16 \AA\ for Ni.
The potential is calculated self-consistently with, e. g., for the
calculation of Fig. 3a, 49 \kpara\ points in the two-dimensional Brillouin zone.
Spin-orbit interaction is treated self-consistently as described
in Ref. [\onlinecite{Li}].

For simplicity, the Au monolayer
was assumed to be pseudomorphic [$p(1\times1)$] to the graphene (graphene lattice
constant 2.489 \AA) and to the Ni(111) substrate with the same area density of Ni(111).
This is also the basis for the 0.25 ML Au in the $p(2\times2)$ structure.
The important Au-graphene distance has been chosen in Fig. 3a based on the Fermi-level position
and in agreement with Ref. [\onlinecite{Giovannetti-PRL-2008}] as 3.3 \AA. The dependence
of the Fermi level and spin-orbit splitting on this parameter is shown  
in Fig. S10. For the calculation shown in Fig. 3c a structural optimization has been conducted 
by force minimization.

Acknowledgement: 
E. I. R. acknowledges funding from the NSF Materials World Network and the 
Intelligence Advanced Research Project Activity (IARPA) through the Army Research Office.
This work was supported by SPP 1459 of the Deutsche Forschungsgemeinschaft.

\textcolor{white}{
\section*{Figures}
}
\addtocounter{section}{1}
\phantom{Zeile}
\begin{figure}[h!]
\includegraphics[width=1\textwidth]{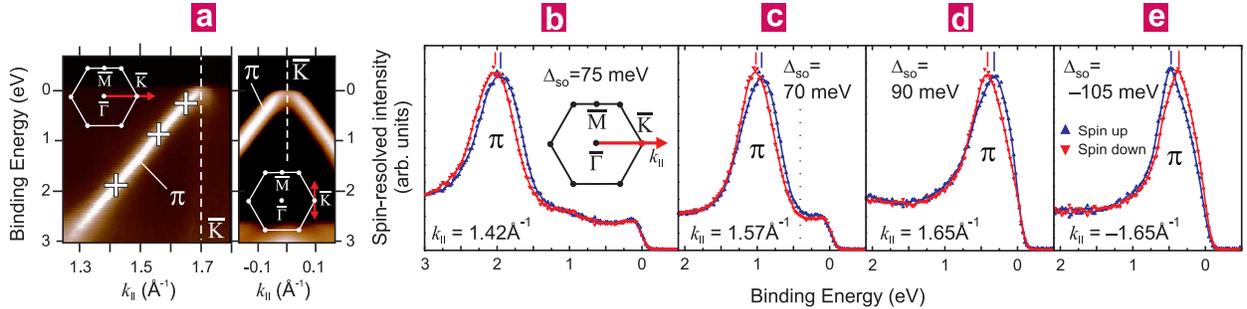}
\caption{
{\bf Giant spin-orbit splitting $\Delta_{\rm so}$ of the graphene $\pi$ band.}
(a) Angle-resolved photoemission of graphene/Au/Ni(111) with indication by crosses where the Dirac cone of graphene $\pi$-states is probed by
(b--e) spin- and angle-resolved photoemission spectra ($h\nu=50$ eV). The spin splitting reverses with the sign of \kpara\ as expected for a Rashba effect.
}
\label{Fig1}
\end{figure}

\phantom{Zeile}
\begin{figure}[h!]
\includegraphics[width=1\textwidth]{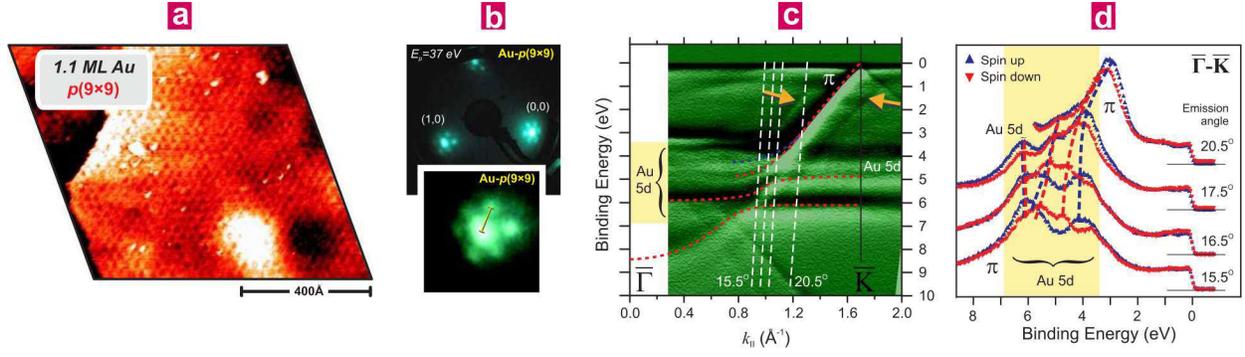}
\caption{
{\bf Geometrical structure and hybridization of the intercalated Au monolayer under graphene.}
(a) Scanning tunneling microscopy and (b) low-energy electron diffraction show  a \ninebynine\ moir\'e-type superstructure. 
(c) Angle-resolved photoemission   reveals hybridization of graphene $\pi$-bands and Au $d$ states. Moreover,  $\pi$-bands shifted to larger and smaller  \kpara\ appear due to the \ninebynine\ superstructure (orange arrows). Dashed lines in (c) show at which \kpara\ points spin-dependent spectra are displayed in (d) ($h\nu=62$ eV). It is seen that the spin-split $\pi$ states develop directly out of a large spin-orbit splitting of Au$5d$ states
}
\label{Fig2}
\end{figure}

\phantom{Zeile}
\begin{figure}[h!]
\includegraphics[width=1\textwidth]{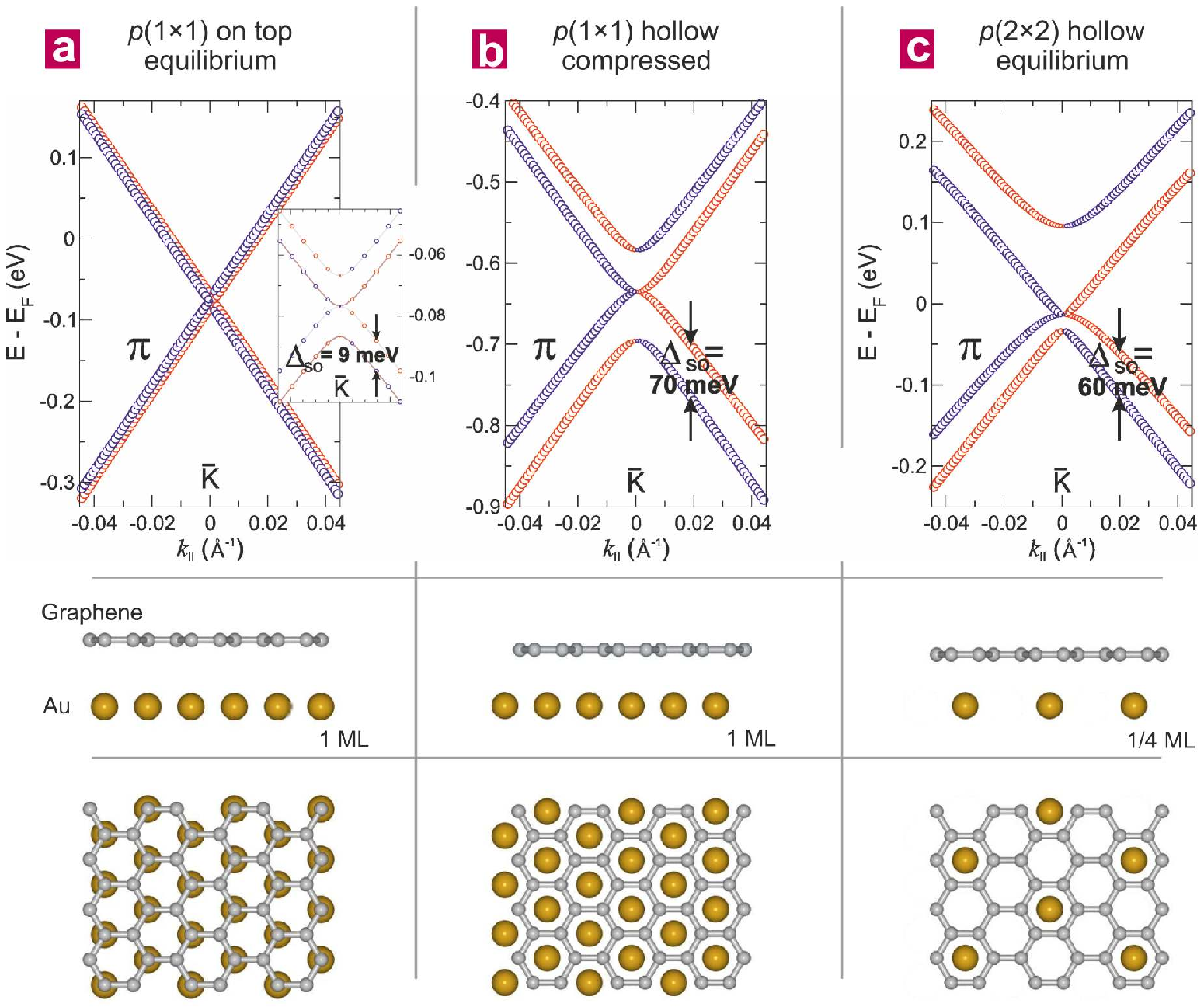}
\caption{
{\bf Interface geometry and Dirac-cone splitting.}
(a) {\it Ab initio} calculation showing the Rashba-split Dirac cone for graphene and a Au monolayer in the on-top position. The approximate equilibrium distance of 3.3 \AA\ leads to a spin-orbit splitting of 9 meV at the Fermi level.  
(b) The Au monolayer is   laterally moved to the graphene hollow sites and can now be pressed into the graphene to the non-equilibrium distance of 2.5 \AA\ without breaking the Dirac cone. This leads to a giant  spin-orbit splitting of $\sim70$ meV. 
(c) Improved model with the Au atoms still in the graphene hollow sites but diluted to 0.25 ML Au in a $p(2\times2)$ geometry. This lowers the equilibrium separation between the graphene and the Au layer to about $2.3$ \AA\ and causes a Fermi-level position similar to the experiment and a giant ($50$ to $100$ meV) spin-orbit splitting.
}
\label{Fig3}
\end{figure}

 \phantom{x}
\newpage
 \phantom{x}
\newpage
 \phantom{x}



\textcolor{white}{
\section*{Supplementary Information}
}
\addtocounter{section}{1}

\renewcommand{\figurename}{Fig. S}
\setcounter{figure}{0}

 \newpage
 \phantom{x}

\noindent {\bf SUPPLEMENTARY INFORMATION}

\begin{center}
\textbf{Graphene for spintronics: giant Rashba splitting due to hybridization with Au}
\end{center}

\noindent{D. Marchenko, A. Varykhalov, M. R. Scholz, G. Bihlmayer, E. I. Rashba, A. Rybkin, A. M.
Shikin, O. Rader}

\begin{figure}[h!]
\includegraphics[width=1\textwidth]{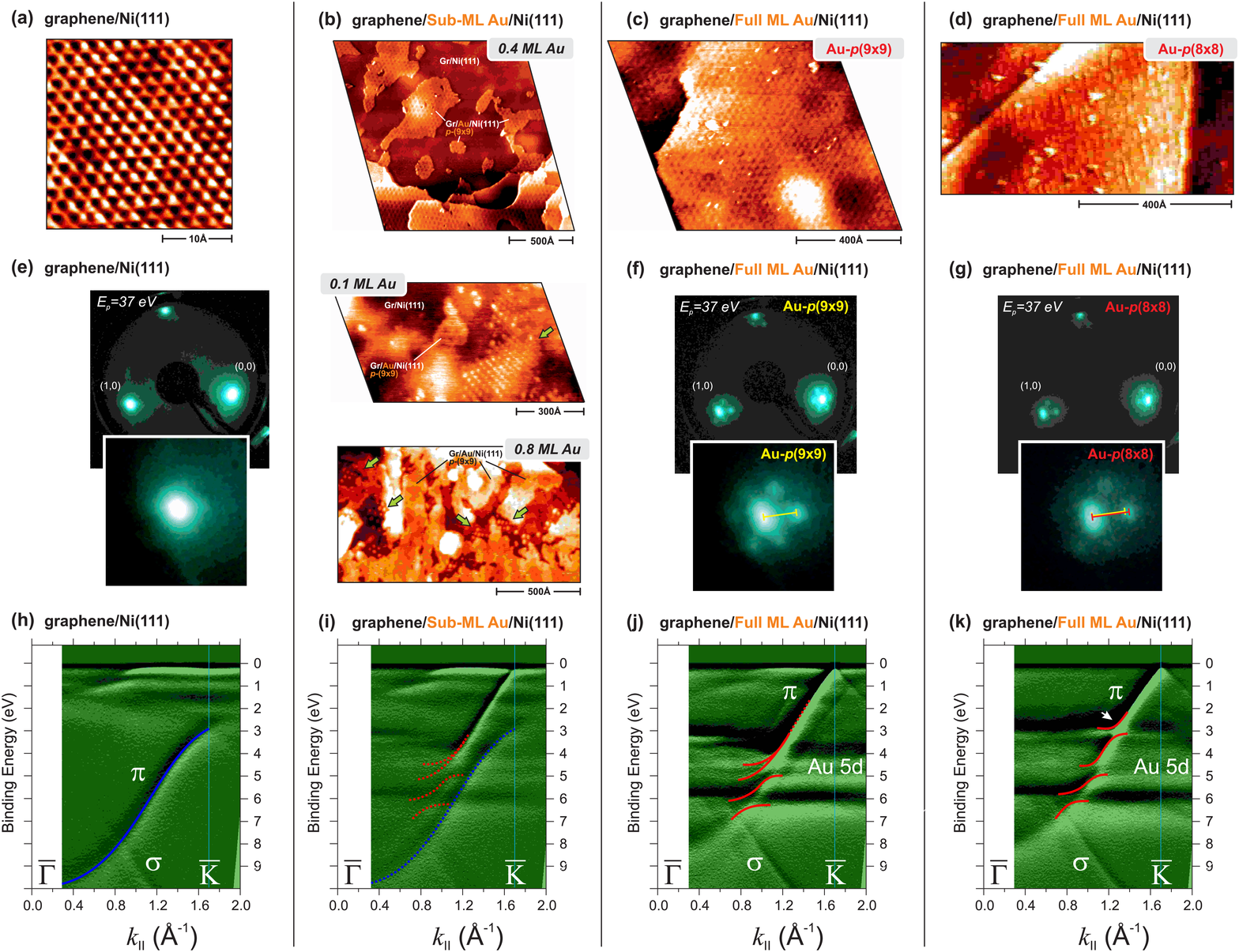}
\caption{
\textbf{Structural characterization of the monolayer graphene/Au phase.} 
}
\label{FigS1}
\end{figure}
We characterize graphene on Ni(111) intercalated with different amounts of Au by means of scanning tunneling microscopy (STM) and low-energy electron diffraction (LEED): (a,e) Graphene on bare Ni(111) shows a pronounced 3-fold symmetry in STM and a clear $p(1 \times 1)$ pattern in LEED which means that graphene is perfectly in registry to the Ni substrate. (b) Graphene on Ni(111) intercalated with various submonolayer amounts of Au. Underneath of the graphene, the Au forms islands of various dimensions and shapes. STM scans of these islands exhibit the $p(9 \times 9)$ moir\'e superstucture. This superstructure is attributed to lattice mismatch between Au and Ni. The interatomic distances are in bulk Au 2.88 {\AA} and in bulk Ni 2.48 {\AA}. Since the lattices of graphene and Ni(111) match exactly, it is not surprising that 1 ML Au/Ni(111) also forms a $p(9 \times 9)$ structure at room temperature [Jac95, Ume98]. Underneath the graphene, this structure is rather stable since independently of the exact amount of intercalated Au, these $p(9 \times 9)$ islands always co-exist with areas which resemble cluster superlattices (marked with green arrows). Those areas can be attributed to the formation of an interfacial alloy between Ni and Au under the graphene. Such surface alloy has been observed for Au/Ni(111) after annealing [Ves05]. However, these clusters cover only a minor part of the sample surface and were not found relevant to our results. The fact that we do not observe large areas of alloyed Au under the graphene is understandable: It was shown and demonstrated for CO that the presence of another species reverses the process again (de-alloying) [Ves05]. (c,f) Graphene on Ni intercalated with a full Au monolayer (nominally 1.1 ML) demonstrates in STM and LEED a perfectly periodic moir\'e pattern. A quantitative analysis is easier for the LEED than for the STM since the superstructure can be evaluated relative to the distance between (0,0) and (1,0) spots without the need for calibration. Our data reveals that the moir\'e is due to a $p(9 \times 9)$ superstructure. (d,g) Further increase of the Au amount leads to a $p(8 \times 8)$ structure which is less ordered in STM but clearly distinguishable in LEED. When the Au is deposited as a wedge, scanning the sample during LEED shows a clear jump between the $p(9 \times 9)$ and the $p(8 \times 8)$. (h-k) Corresponding development of the hybridization of graphene with Au states from a wedge-type sample. The $p(8 \times 8)$ phase leads to an additional hybridization around 3 eV binding energy. The SARPES studies of the giant Rashba splitting presented here refer to the $p(9 \times 9)$ phase but the Rashba splitting appears similar in the $p(8 \times 8)$ phase.
\newpage

\begin{figure}[H]
\includegraphics[width=1\textwidth]{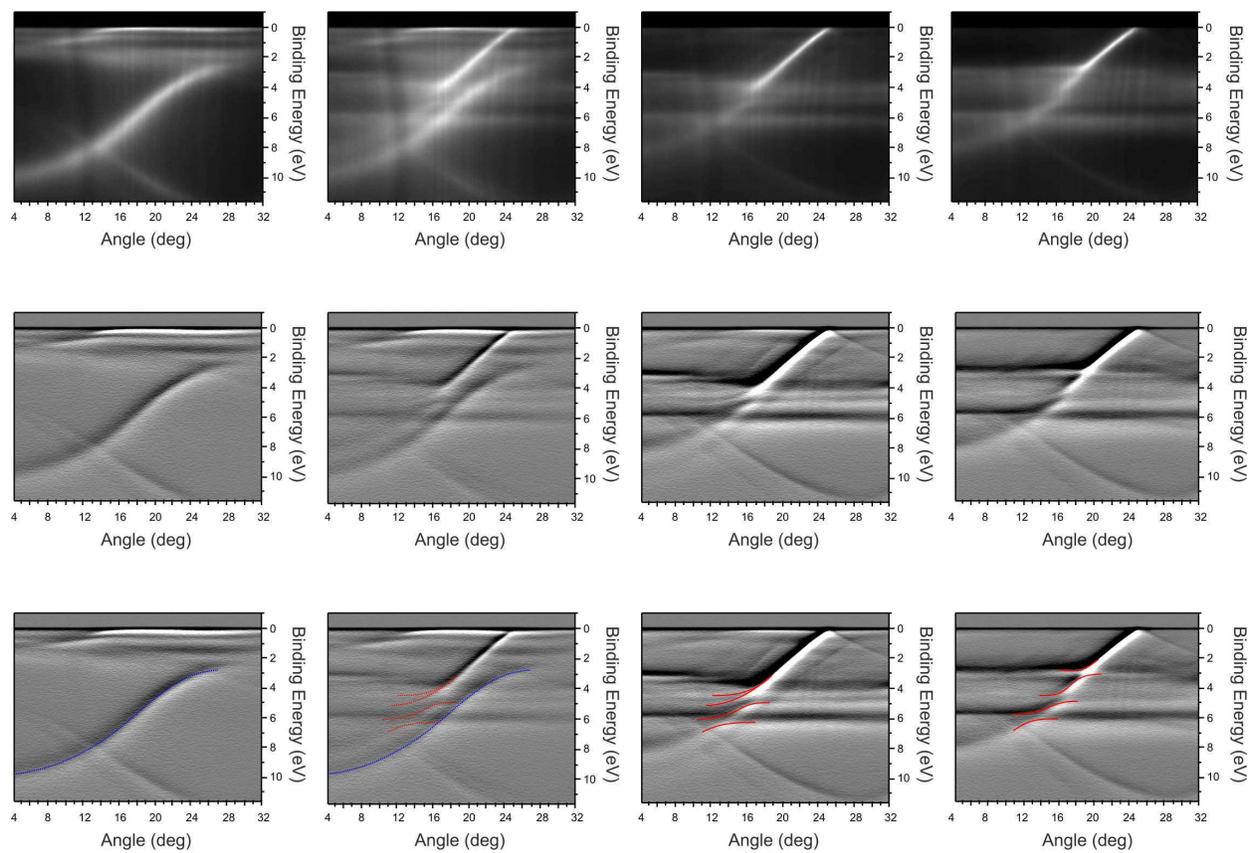}
\caption{
\textbf{Original data of Fig. S1.} Intensity (top) and first derivative without (middle) and with marks (bottom).
}
\label{FigS2}
\end{figure}

\begin{figure}[H]
\includegraphics[width=0.8\textwidth]{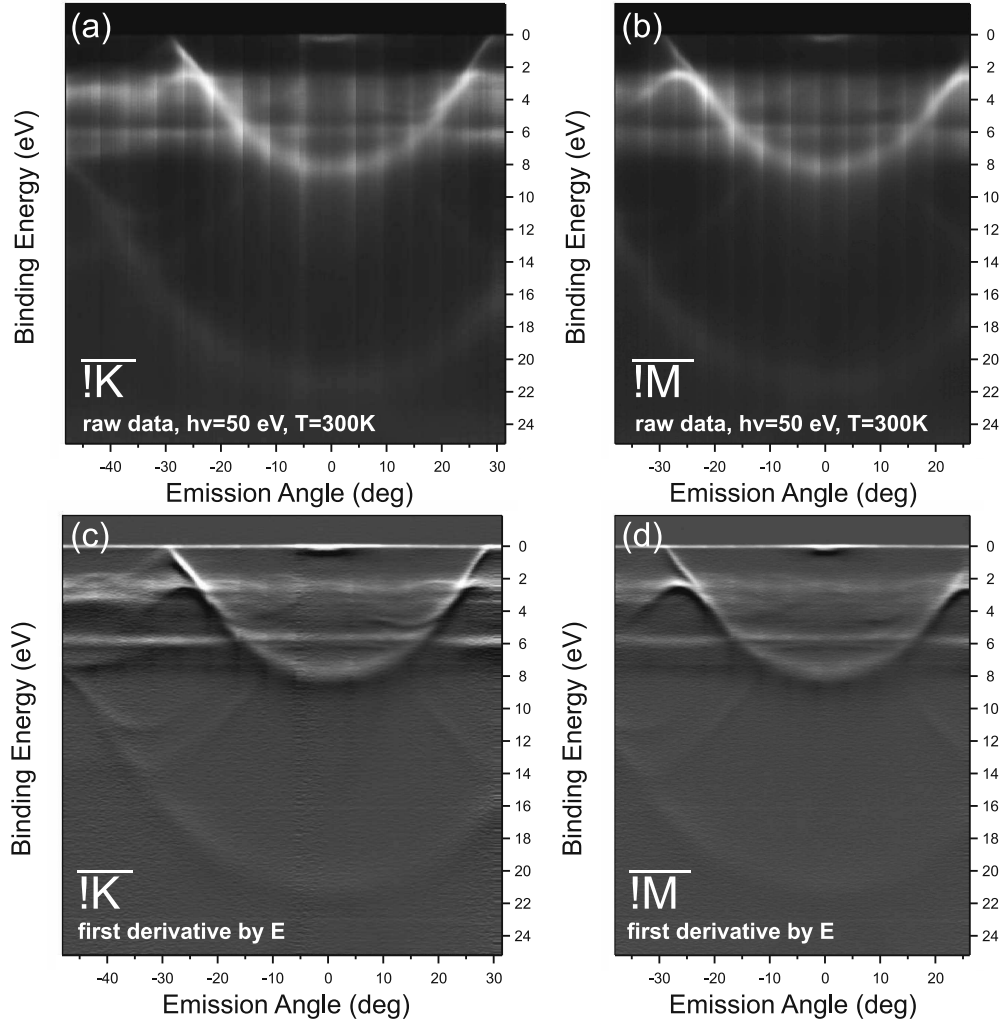}
\caption{
\textbf{Comparison to the data of [Var08a].} Previously we observed a Rashba splitting of ~13 meV for graphene/Au/Ni(111) [Var08a]. Fig. S3 shows published and unpublished data of the measurements of [Var08a]. Apparently, the dispersions do not look similar to any of the wedge sample of Fig. S1. The main reason is the appearance of extra \Gbar\Mbar dispersions (no Dirac cone) along \Gbar\Kbar (a,c). Similarly the measurements (b,d) along \Gbar\Mbar (which should not show a Dirac cone) contain contributions from \Gbar\Kbar. This is due to rotational domains. The presence of domains means that the graphene layer was imperfect. This does not necessarily affect the SARPES measurement of the p-band in the vicinity of the Dirac point but spectroscopically the hybridization gaps seen clearly in the present data are superimposed with spurious band dispersions. Most importantly, the presence of domain boundaries facilitates the intercalation of the Au since imperfections of the graphene are considered the main pathway for intercalation of deposited metals [Ton91]. This leads to a different structure of the intercalated Au. The present data in Fig. S1(j,k) do not show such rotational domains any more.
}
\label{FigS3}
\end{figure}

\begin{figure}[H]
\includegraphics[width=1\textwidth]{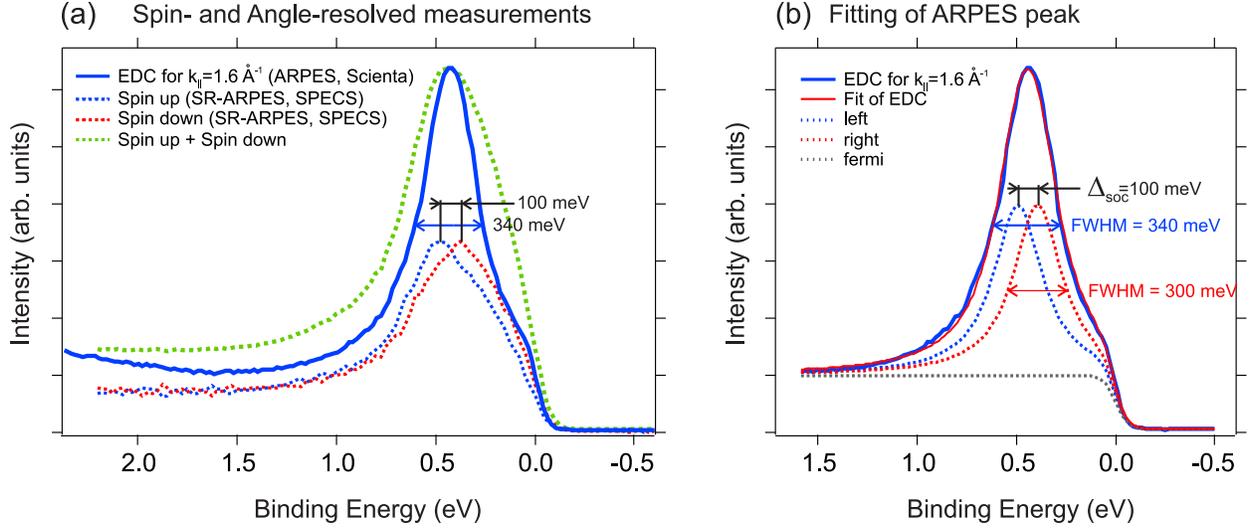}
\caption{
\textbf{Consistency of the spin-resolved and non-spin-resolved ARPES measurements.} The addition of the spin-up and the spin-down SARPES spectrum leads to a broad peak (dashed green) in (a). On the other hand, the ARPES data [solid blue in (a)] is sharper due to higher angle resolution of the ARPES setup. Therefore, in (b) we repeated the addition of two peaks split by 100 meV. It is seen that also in this case no splitting appears after adding the spectra.
}
\label{FigS4}
\end{figure}

\begin{figure}[H]
\includegraphics[width=0.8\textwidth]{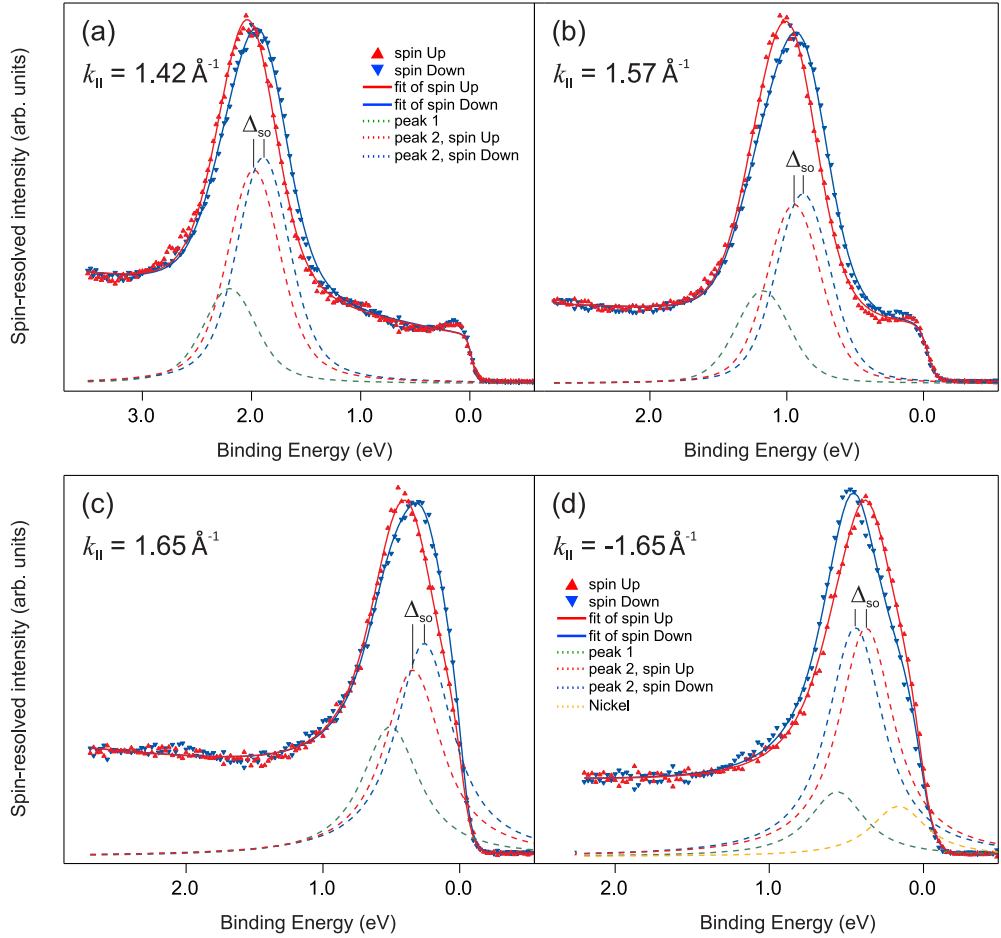}
\caption{
\textbf{Contributions of the phase with low spin orbit splitting to the spin-resolved spectra.} The spectra show a rather asymmetric splitting suggesting a less or non-split contribution at higher binding energy. Such phase would correspond to the one with $\approx$13 meV spin-orbit splitting in [Var08a]. The figure shows the attempt to fit the spectra under the most simple assumption: In addition to the spin-up and spin-down components (here at lower binding energy) we allow for a non-split component (as an approximation to the small splitting of 13 meV). Because the 4 spectra stem from 3 measurements with slightly different ratios between the coexisting phases, their ratios are allowed to vary in the fit as well. The negative wave vector -1.65 {\AA} can only be reached by a large change in the electron emission angle. This changes also the light polarization conditions. Apparently, an additional component for Ni 3d states at the Fermi energy had to be introduced in this case.
}
\label{FigS5}
\end{figure}

\begin{figure}[H]
\includegraphics[width=1\textwidth]{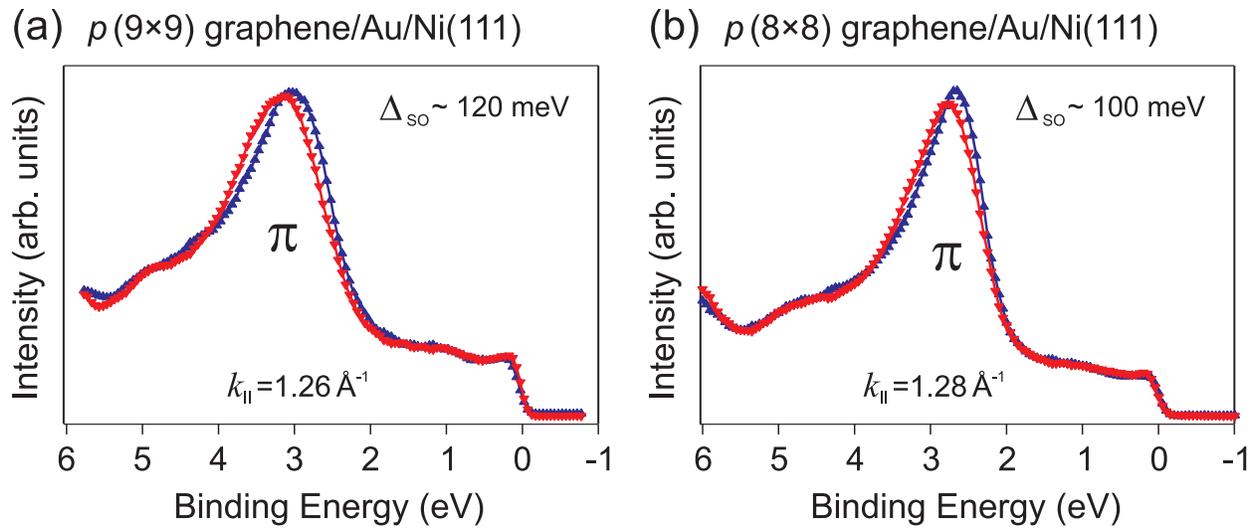}
\caption{
\textbf{Comparison of spin-orbit splitting of $p(9 \times 9)$ and $p(8 \times 8)$ structure.} Spin-resolved spectra are compared from a $p(9 \times 9)$ and a predominant $p(8 \times 8)$ superstructure. The spin splitting is very similar.
}
\label{FigS6}
\end{figure}

\begin{figure}[H]
\includegraphics[width=1\textwidth]{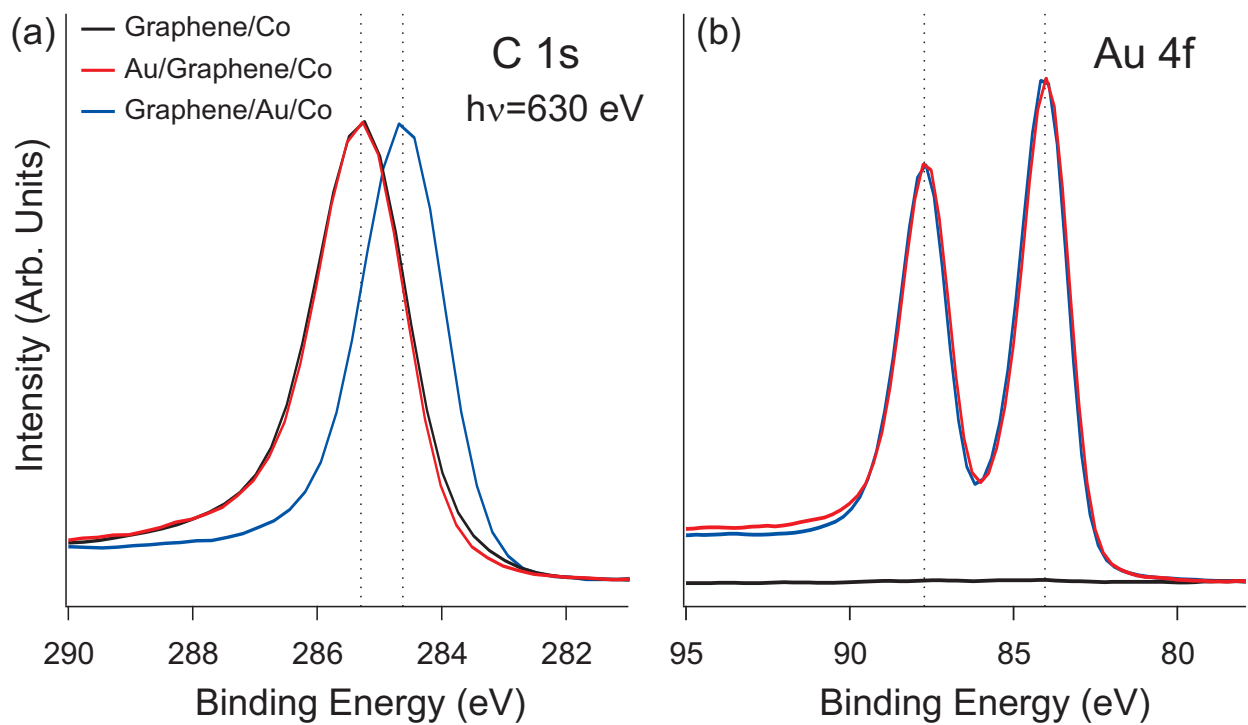}
\caption{
\textbf{Au 4f core-level spectra of the intercalation process.} The spectra before and after intercalation do not show a core-level shift which could be used to identify the sites of Au atoms. After graphene growth (black) the Au ($\approx$1.5 ML) is deposited at room temperature (red) and subsequently intercalated by annealing (blue). The C1s shift shows the successful intercalation whereas the Au4f does not shift visibly. The data are for Co(0001) instead of Ni(111) but otherwise comparable. The spectra have been normalized to equal maximum intensity.
}
\label{FigS7}
\end{figure}

\begin{figure}[H]
\includegraphics[width=0.6\textwidth]{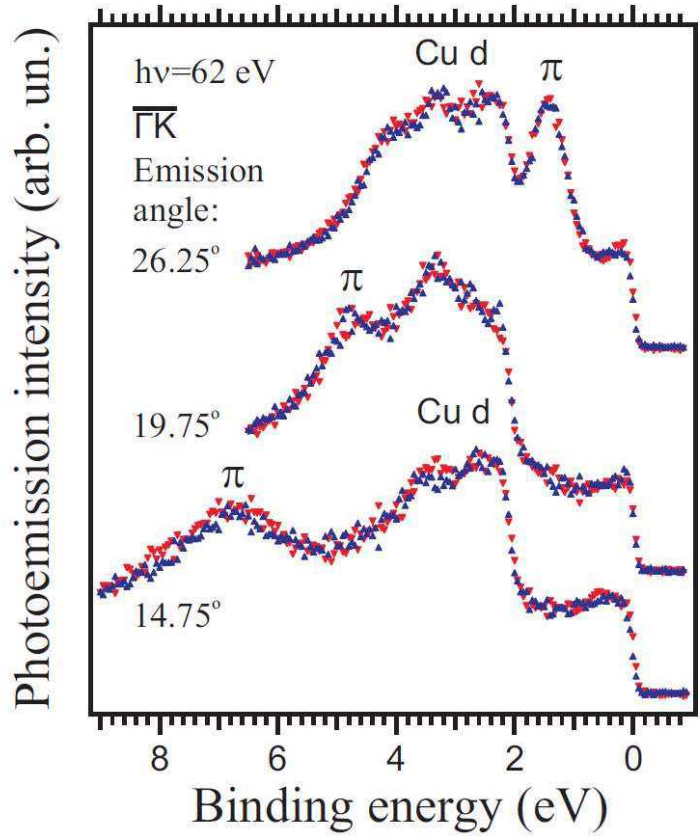}
\caption{
\textbf{Cu-intercalated graphene\/Ni(111) as control sample.} After intercalation of a monolayer Cu instead of Au we do not observe a any substantial spin splitting neither in the graphene $\pi$-band nor in the Cu3d states themselves. Colors distinguish spin up and spin down. The geometry is comparable to Fig. 1 in the manuscript.
}
\label{FigS8}
\end{figure}

\begin{figure}[H]
\includegraphics[width=0.4\textwidth]{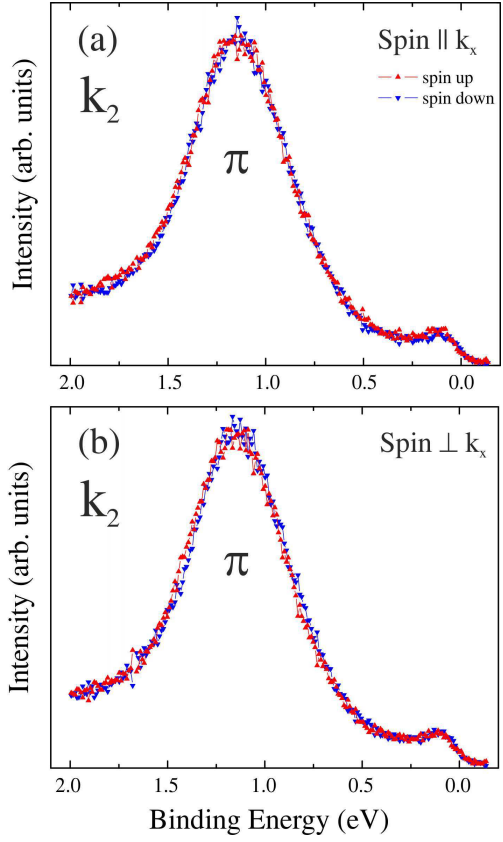}
\caption{
\textbf{Graphene grown on SiC as control sample.} Spectra for two orthogonal spin directions in the surface plane. No spin splitting is seen, and the upper limit for the spin splitting from this measurement is $<9$ meV parallel to $k_x$ and $<15$ meV for a splitting parallel to $k_y$.
}
\label{FigS9}
\end{figure}

\begin{figure}[H]
\includegraphics[width=0.8\textwidth]{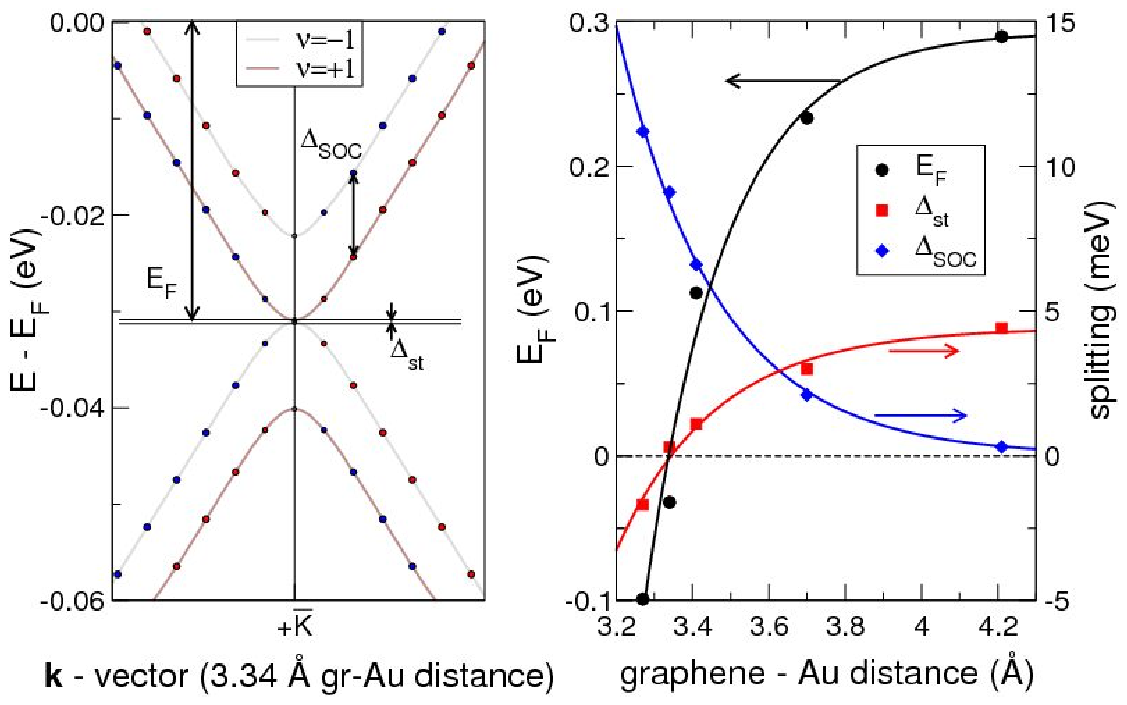}
\caption{
\textbf{Band topology and distance dependence of doping and spin-orbit splitting.} Left: We compare the analytically calculated Rashba-type band topology (solid lines) to the calculation of Fig. 3(a) (symbols) [$p(1 \times 1)$ on-top geometry] in the vicinity of the Dirac point and find full agreement. The line color gives the chirality n and the symbol color (red and blue) the spin polarization. The spin-orbit splitting $\Delta_{SOC}$ amounts to 9 meV. For this structure we varied the graphene-Au distance. Right: The spin-orbit splitting strongly depends on the distance of the graphene to the heavy Au atoms. This finding is in line with the results from pure metal surfaces such as Au(111) and Ag(111) where the surface and nuclear potentials were found to contribute multiplicatively [Pet00, Bih06, Nag09]. The sign change from n- to p-doping between d = 3.34 and 3.41 {\AA} is in agreement with calculations for R-($\sqrt{3} \times \sqrt{3}$) graphene/Au(111) which give an equilibrium graphene-Au distance of 3.31 {\AA} [Gio08]. In addition, the behavior of the staggered potential ($\Delta_{st}$) is shown. The staggered potential is induced by the Au lattice which breaks the equivalence of A and B sublattices of the graphene if one sublattice is in the on-top position relative to the Au as shown in the geometry sketch of Fig. 3(a). This effect from the staggered potential causes a splitting of the Dirac cone of a few meV for large graphene-Au distances but it increases strongly for short distances breaking the Dirac cone.
}
\label{FigS10}
\end{figure}

\begin{figure}[H]
\includegraphics[width=0.5\textwidth]{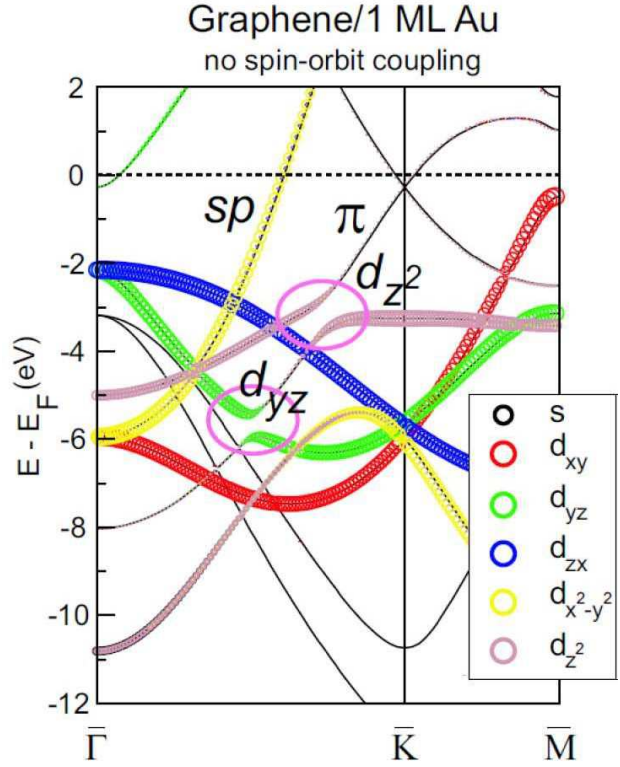}
\caption{
\textbf{Hybridization as the origin of the spin-orbit splitting in the graphene.} Calculations without spin-orbit interaction reveal anticrossings and hybridization gaps which are due to orbital symmetries. (The spin-orbit splitting lowers the symmetry further and introduces additional anticrossings.) The model is $p(1 \times 1)$ graphene on a Au monolayer in the on top geometry. Since the $\pi$-band is made of $p_z$ orbitals, the observed hybridization with $d_z^2$ and $d_{yz}$ orbitals is determined by symmetry.
}
\label{FigS11}
\end{figure}

\begin{figure}[H]
\includegraphics[width=0.5\textwidth]{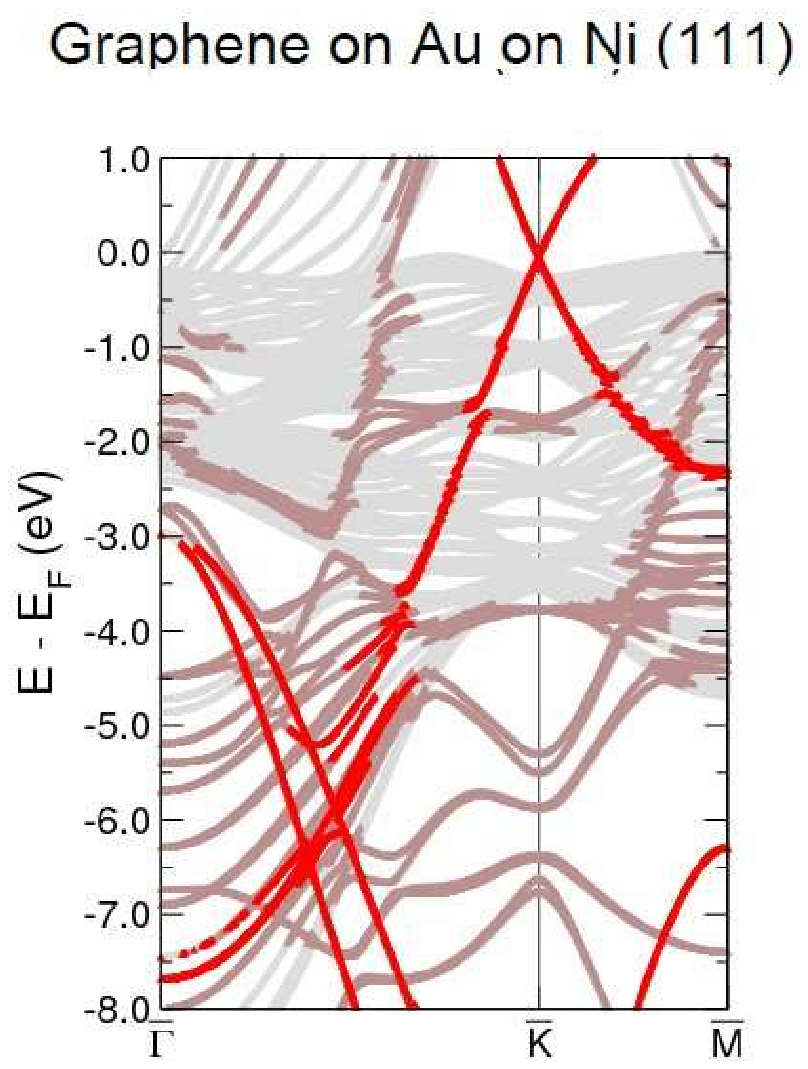}
\caption{
\textbf{Effect of the Ni substrate on the electronic structure at the graphene/Au interface.} The spin-orbit splitting in the graphene clearly stems from the contact with the heavy Au atoms and we restricted the modeling in the manuscript to this bilayer of graphene and Au. Nevertheless, we have also investigated which principle influence the Ni substrate underneath has. Colors mark states with a high probability at the graphene (red), Au monolayer (violet), and Ni substrate (grey). The experimental lattice constant of Ni(111) agrees within 1\% to that of graphene and was chosen to be equal here. The Au monolayer was compressed to the same lattice, as before. The distance graphene-Au was set to 3.3 {\AA}. This distance nicely reproduces the position of the graphene states from the experiment. The same distance was used in the graphene-Au bilayer model of Fig. S10 but we observe that between these two models the binding energies of the Au5d bands suffer considerable changes due to the Ni substrate, mainly Au-Ni hybridizations which make the Au5d bands difficult to distinguish. The $d_{zx}$ band from Fig. S11 originating at -2 eV at \Gbar\ and arriving at -6 eV at \Kbar\ can be distinguished also here. A new Au5d state appears at -1.5 eV at \Kbar. The Au sp band of Fig. S11 is not observed in the experiment, and the present figure shows that the reason is that this band is very strongly hybridized with the Ni and disappears as a well-defined band.
}
\label{FigS12}
\end{figure}

\begin{figure}[H]
\includegraphics[width=1\textwidth]{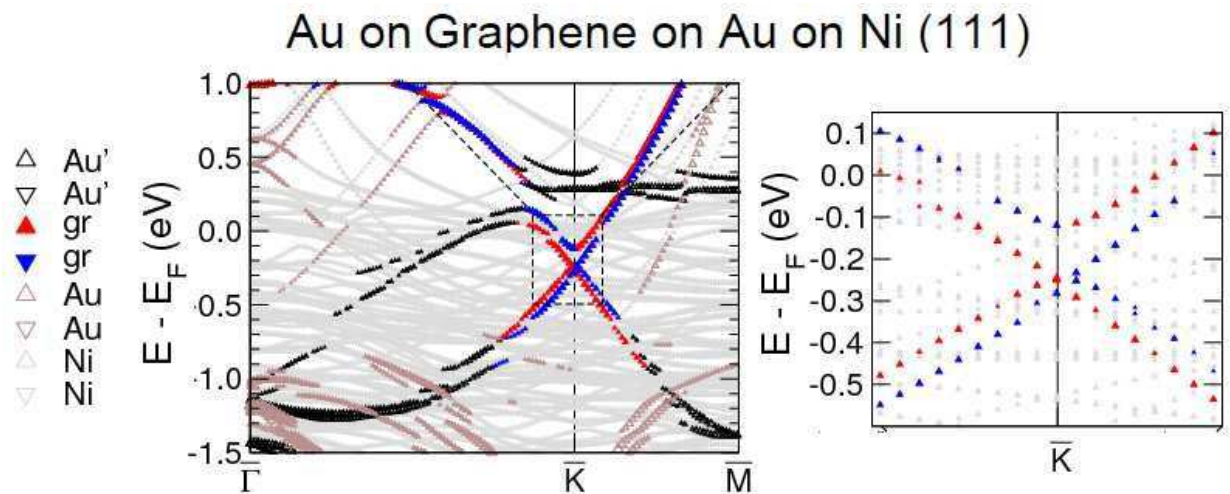}
\caption{
\textbf{Same model as in Fig. S12 but with additional Au adatoms.} This calculation explores again the influence of the Ni substrate (grey symbols) under the Au monolayer (violet symbols) on the graphene $\pi$-states (red and blue symbols). In this case, a giant Rashba splitting similar to Fig. 3(c) is achieved by adding $p(2 \times 2)$ 0.25 ML Au as adatoms on the surface in graphene hollow sites in the geometry of Fig 3c. In this way the graphene is sandwiched between 0.25 ML and 1 ML Au. The upward and downward direction of the triangles marks the spin, in particular the upward red and downward blue triangles for graphene states. Left: The states due to the additional Au adatoms (black triangles, Au') can be seen to interact with the graphene states through different hybridizations than the Au monolayer. Right: The spin-orbit split Dirac cone from the graphene-Au bilayer model of Fig. 1c is thus confirmed in the present, more realistic, configuration. The fact that higher and lower energies are not symmetric about the Dirac point is not due to the Ni substrate but to the $p(2 \times 2)$ 0.25 ML Au adatoms, like in Fig. 3(c) of the manuscript.
}
\label{FigS13}
\end{figure}

\begin{figure}[H]
\includegraphics[width=0.5\textwidth]{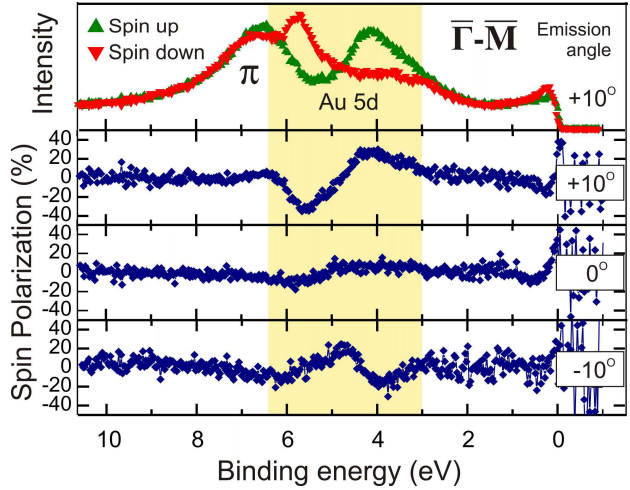}
\caption{
\textbf{Origin of the spin-orbit splitting in the Au.} Comparison of $+k_{\parallel} = 0.6$ {\AA} and $-k_{\parallel}$ data for graphene/Au/Ni(111). The behavior of the spin polarization [$p = (I_\uparrow-I_\downarrow)/(I_\uparrow+I_\downarrow)$] shows an almost vanishing spin polarization at $k_{\parallel} = 0$, as expected for a Rashba effect, and a rather clear reversal for $-k_{\parallel}$ if we take into account that the spin polarization of Au5d states is also subject to spectroscopic effects such as the linear dichroism of Au5d emission [Sto91] which do not reverse with the sign of $k_{\parallel}$. This means that the Rashba-type spin-orbit splitting of the graphene $\pi$ states is accompanied by a Rashba effect on the Au monolayer itself. Such effect has similarly been observed for Au/W(110) [Shi08, Var08b].
}
\label{FigS14}
\end{figure}

\textcolor{white}{
\section*{References of Supplementary Information}
}
\addtocounter{section}{1}

 \phantom{x}

\noindent {\bf References}

\noindent [Bih06] G. Bihlmayer, Yu. M. Koroteev, P. M. Echenique, E. V. Chulkov and S. Bl\"ugel, The Rashba effect at metallic surfaces, Surf. Sci. 600, 3888 (2006).

\noindent [Gio08] G. Giovannetti, P. A. Khomyakov, G. Brocks, V. M. Karpan, J. van den Brink, and P. J. Kelly, Doping graphene with metal contacts, Phys. Rev. Lett. 101, 026803 (2008).

\noindent [Jac98] J. Jacobsen, L. P. Nielsen, F. Besenbacher, I. Stensgaard, E. L{\ae}gsgaard, T. Ramussen, K. W. Jacobsen, and J. K. N{\o}rskov, Atomic-Scale Determination of Misfit Dislocation Loops at Metal-Metal Interfaces, Phys. Rev. Lett. 75, 489 (1995).

\noindent [Nag09] M. Nagano, A. Kodama, T. Shishidou, and T. Oguchi, A first-principles study on the Rashba effect in surface systems, J. Phys.: Condens. Matter 21, 064239 (2009).

\noindent [Pet00] L. Petersen and P. Hedegård, A simple tight-binding model of spin-orbit splitting of sp-derived surface states, Surf. Sci. 459, 49 (2000).

\noindent [Shi08] A. M. Shikin, A. Varykhalov, G. V. Prudnikova, D. Usachov, V. K. Adamchuk, Y. Yamada, J. D. Riley, and O. Rader, Origin of Spin-Orbit Splitting for Monolayers of Au and Ag on W(110) and Mo(110), Phys. Rev. Lett. 100, 057601 (2008).

\noindent [Sto91] P. Stoppmanns, B. Heidemann, N. Irmer, N. Müller, B. Vogt, B. Schmiedeskamp, U. Heinzmann, E. Tamura, and R. Feder, Au induced surface state on Pt(111) revealed by spin resolved photoemission with linearly polarized light, Phys. Rev. Lett. 66, 2645 (1991).

\noindent [Ton91] A. Ya. Tontegode, Carbon on transition metal surfaces, Prog. Surf. Sci. 38, 201 (1991).

\noindent [Ume98] K. Umezawa, S. Nakanishi, and W. M. Gibson, Growth modes depending on the growing temperature in heteroepitaxy: Au/Ni(111), Phys. Rev. B 57, 8842 (1998).

\noindent [Var08a] A. Varykhalov, J. S\'anchez-Barriga, A. M. Shikin, C. Biswas, E. Vescovo, A. Rybkin, D. Marchenko, and O. Rader, Electronic and magnetic properties of quasifreestanding graphene on Ni, Phys. Rev. Lett. 101, 157601 (2008).

\noindent [Var08b] A. Varykhalov, J. S\'anchez-Barriga, A. M. Shikin, W. Gudat, W. Eberhardt, and O. Rader, Quantum cavity for spin due to spin-orbit interaction at a metal boundary, Phys. Rev. Lett. 101, 256601 (2008).

\noindent [Ves05] E. K. Vestergaard, R. T. Vang, J. Knudsen, T. M. Pedersen, T. An, E. L{\ae}gsgaard, I. Stensgaard, B. Hammer, and F. Besenbacher, Adsorbate-Induced Alloy Phase Separation: A Direct View by High-Pressure Scanning Tunneling Microscopy, Phys. Rev. Lett. 95, 126101 (2005).

\end{document}